\def\den{\hbox{den}}

\def\tr{\hbox{tr}}
\def\ln{\ell{n}}
\documentstyle[12pt]{article}

 {\parindent 0.5cm \textwidth 15.0cm \textheight
23.5cm \hoffset=-0.8cm \voffset=-3cm
  \let\LARGE=\large
 \let\large=\normalsize
\newcommand{\be}{\begin{equation}}
\newcommand{\ee}{\end{equation}}
\newcommand{\ba}{\begin{array}{c}}
\newcommand{\ea}{\end{array}}

\begin{document}
\begin{titlepage} \vspace{0.2in} \begin{flushright}
MITH-95/6 \\ \end{flushright} \vspace*{1.5cm}
\begin{center} {\LARGE \bf  The Emergence of a Heavy Quark Family on a Lattice 
\\} \vspace*{0.8cm}
{\bf Giuliano Preparata and She-Sheng Xue$^{a)}$}\\ \vspace*{1cm}
INFN - Section of Milan, Via Celoria 16, Milan, Italy\\ \vspace*{1.8cm}
{\bf   Abstract  \\ } \end{center} \indent
 
Within the framework of the ``Rome approach'' for a lattice chiral gauge theory,
the four-quark interaction with flavour symmetry is included. We analyse 
spontaneous symmetry breaking and compute composite modes and their contributions
to the ground state energy. As a result, it is shown that the emergence of a 
heavy quark family is the energetically favoured solution.

\vfill \begin{flushleft}  June, 1995 \\
PACS 11.15Ha, 11.30.Rd, 11.30.Qc  \vspace*{3cm} \\
\noindent{\rule[-.3cm]{5cm}{.02cm}} \\
\vspace*{0.2cm} \hspace*{0.5cm} ${}^{a)}$ 
E-mail address: xue@milano.infn.it\end{flushleft} \end{titlepage}
 
\noindent
{\bf 1.}\hspace*{0.3cm} 
The problem of how mass gets generated in the Standard Model (SM) of
fundamental interactions, $SU_c(3)\otimes SU_L(2)\otimes U_Y(1)$, is perhaps
the most important that is now facing both theoretical and experimental
high-energy physics. As well known, the mechanism of mass generation now
generally considered, based on the Anderson-Higgs-Kibble mechanism associated
to a fundamental local scalar isodoublet Yukawa-coupled to the basic
Fermi-fields (quarks and leptons), is also generally believed to be ``too ugly"
to be really fundamental, leading to the conviction that it must be but the
simple surrogate of a deeper, yet to be found and understood, layer of particle
interactions. Of particular interest in this direction is the
phenomenologically proposed $\bar tt$-condensate model\cite{nambu1}\cite{bar},
in the light of the experimental observation that the quark family ($t,b$) is
much heavier than others. This model revives the Nambu-Jona~Lasinio (NJL)
proposal\cite{nambu2} of a 4-fermion interaction. However it cannot be denied 
that the addition of
an NJL-interaction of the quark family $(t, b)$ only to the usual
gauge-invariant Lagrangian density gravely lacks compelling motivation. In
this paper, it is shown that the $\bar t t$-condensate model emerges
as an energetically favoured solution in a lattice-regularized SM with the
extension of a four-quark interaction. 

The fermion ``doubling'' phenomenon is a well-known problem arising when
fermion fields are defined on a lattice. In fact, the ``no-go" theorem of
Nielsen-Ninomiya\cite{nogo}, which stipulates that no simple ``lattice
transcription" of the bilinear fermion Lagrangian of SM exists, indicates that
the SM on a lattice may include extra gauge-symmetric quardrilinear
interactions ($S_4$)\footnote{Here we are not claiming to solve the problem of
chiral gauge theories on a lattice by adding quardrilinear terms}\cite{ep}\cite{xue91}. Thus we ask whether
the physical incompleteness (lack of mass-generation) of the SM, as formulated
in continuous space-time, could not be the symptom of a basic lattice structure
of space-time: the arena of physical reality\cite{wheeler}. 

In order to remove doublers, for each quark we add Wilson terms ($S_w$)
\cite{wilson} that explicitly breaks chiral gauge symmetries of the SM. For
the purpose of obtaining the SM in the low-energy region (target theory), we
adopt the ``Rome approach''\cite{rome} by adding all necessary counterterms
($S_{ct}$) to allow tuning to impose the satisfying of the Ward identities
associated with gauge symmetries of the SM\footnote{Since we shall consider
only computation of gauge invariant quantities, gauge fixing and ghosts fields
and BRST symmetry are not introduced.}. Thus, we have the following Lagrangian
for the quark sector:
\begin{equation}
S=S_d+S_4+S_w+S_{ct},
\label{action}
\end{equation}
where $S_d$ is the naive lattice transcription of the SM. The Wilson term and 
the four-quark interaction are
\begin{eqnarray}
S_w&=&{r\over a}\sum_{x,\mu}\bar\psi(x)\Big(\psi(x+a_\mu)+
\psi(x-a_\mu)-2\psi(x)\Big);\\
S_4&=&-G_1\sum_x\bar\psi^i_L(x)\cdot\psi^j_R(x)
\bar\psi^j_R(x)\cdot\psi^i_L(x),
\label{sw}
\label{s4}
\end{eqnarray}
where ``$a$'' is the lattice spacing and ``$i,j$'' are indices of quark family
and weak isospin. 

Due to the fact that the gauge-variant regulator (the Wilson term $S_w$) is
compensated by the gauge-variant counterterms $(S_{ct})$ by forcing satisfying
of Ward identities associated with gauge symmetries of the SM, the total action,
eq.(\ref{action}), is symmetric at the cutoff in the sense that it posseses the
same
symmetries as that of the continuum massless SM. In addition, the total action
(\ref{action}) has global $U(N_g)$ ($N_g$ is the number of quark families)
flavour symmetry and the four-quark interaction $S_4$ introduces the interactions
between quarks in different families. 

In the light of consideration that gauge interactions should not play an
essential role in the mass generation of the heaviest quark family, we
approximately eliminate gauge degree of freedoms in the action (\ref{action}).
It turns out to be a much simpler system of Wilson quarks, the four-quark interaction
and the simplest mass counterterm that must cancel the hard symmetry breaking
term induced by the Wilson term so that the action (\ref{action}) has chiral
symmetry at the cutoff. 

\vspace*{0.5cm}
\noindent
{\bf 2.}\hspace*{0.3cm}
The action (\ref{action}) containing the four-quark interaction ($S_4$),
althought it is forced to satisfy Ward identities at the cutoff, is not
prevented from developing quark mass terms ($m\bar\psi\psi$ dimension-3
operators)\footnote{It should be noted that not only the four-quark interaction
$S_4$ but also Wilson term $(S_w)$ can induce these dimension-3 operators.} that are
soft spontaneous symmetry breakings in the sense that the deviation of imposed
Ward identities is $O(a)$. The massive continuum SM (with $ma\simeq 0, m\not=0$
) should be achieved by careful tunning {\it only} one ``free'' 
parameter\footnote{The ``free'' parameter stands for the parameter free from
being tuned to satisfy Ward identities} in our lattice action (\ref{action})
(that is $G_1$ to be seen later), at the same time, the anomaly of the theory
is restored\cite{smit}.

Since the total action enjoys the flavour and weak
isospin  symmetry, we are allowed to chose a particular basis where the quark
self-energy function $\Sigma^{ij}(p)=\delta^{ij}\Sigma(p)$ is diagonal in the
flavour and weak isospin space. In the planar approximation for the large $N_c$
($N_c\gg 1, N_cG_1$ fixed), one has the 
following Dyson equation for $\Sigma(p)$,
\begin{equation}
\Sigma(p) = -M+{r\over a}w(p)+2g_1 \int_l {\Sigma(l)
 \over \den(l)}
\label{gap0}
\end{equation}
where $g_1a^2 = N_cG_1; l_\mu = q_\mu a, 
\int_l = \int^\pi _{- \pi} {d^4l
\over (2\pi)^4}; w(l)=\sum_\mu(1-\cos l_\mu)$ and 
$\den(l) = \sin^2l_\mu + (a\Sigma(l))^2$. We can write $\Sigma(p)=\Sigma(0)+
{r\over a}w(p)$ and get
\begin{eqnarray}
\Sigma(0) &=& 2g_1 \int_l {\Sigma(0) \over \den(l)};\label{gap1}\\
M &=& 2g_1 \int_l {{ r\over a} w(l) \over \den(l)}.
\label{gap2}
\end{eqnarray}
The first equation is a gap equation of the NJL-type, which has non-trivial 
solution:
$\Sigma(0)\not=0$ for $g_1>g_1^c$(the critical value). The second equation
indicates that the mass counterterm ``$M$'' completely cancel the hard
breaking ``${1\over a}$'' term contributed by the doublers (seeing there is a 
factor $w(l)$ inside the integrand (\ref{gap2})), so as to preserve the chiral 
symmetry at the high energy region. The correspond Ward identity that 
guarantees this cancelation is\cite{rome}, 
\begin{equation}
\langle\psi_L(0)\bar\psi_R(x)\rangle
=0\hskip2cm (x\gg a)
\label{w}
\end{equation}
which must be obeyed up to powers of the lattice spacing $O(a)$. This means
that $a\Sigma(0)$ must be of the order of the lattice spacing $\sim O(a)$ and
gives rise to a soft breaking operator $\Sigma(0)\bar\psi\psi$ that is totally
irrelevant in the high energy region. Thus, we make a consistent fine tuning
on $g_1$ around its critical line $g_1^c(r)$ (Fig.(1)) so that $a\Sigma(0)\sim
O(a)$ at the same time as forcing cancellation (\ref{gap2}) to be obeyed. As a
result, owing to the symmetric action (\ref{action}), we have obtained a
soft spontaneous symmetry breaking $\Sigma(0)=m$ to generate quark mass term
$m\bar\psi\psi$, at the same time doublers are removed from the low energy
spectrum and the anomaly should be reinstated\cite{rome}\cite{smit}. The feature of the
fine tuning of $g_1$, which is very unnatural due to there being no symmetry 
protection, will not be discussed in this paper. 

\vspace*{0.5cm}
\noindent
{\bf 3.}\hspace*{0.3cm}
Composite modes are bound to be produced once the spontaneous
symmetry breakdown occurs $m\not=0$, is evident from the non-trivial solutions
of the gap equation (\ref{gap1}). In order to see these modes, we calculate the
four-quark scattering amplitudes associated with the vertex $S_4$ within the
planar approximation. This
calculation is straightforward and we just present the results. The composite modes
in the pseudo-scalar channel $\Gamma_p(q^2)$ and the scalar channel
$\Gamma_s(q^2)$ are 
\begin{eqnarray}
\Gamma_p(q^2) &=& {{1\over2}} {1 \over I_p(q) A(q^2)},\label{gold}\\
\Gamma_s(q^2) &=& {{1\over2}} {1 \over {4N_c \over a^2} \int_k {[ma +rw(k)]^2 
\over 
D(k,qa)} + I_s(q)A(q^2)}\label{scalar},
\end{eqnarray}
where 
\begin{eqnarray}
A(q^2) &=& \sum_\mu \left({2 \over a} \sin {q_\mu a \over 2}\right)^2,
\nonumber\\
I_p(q) &=& {N_c \over 4} \int_k {c^2 (k) + (r)^2 s^2(k) \over D(k,qa)},
\nonumber\\
I_s(q) &=& {N_c\over 4} \int_k {c^2(k) \over D(k,qa)}\label{i},
\end{eqnarray}
where $c^2(k) = \sum_\mu \cos^2 k_\mu, s^2(k) = \sum_\mu \sin^2 k_\mu$ and
$D(k,qa)=\den (k+{qa\over 2})\den (k-{qa\over 2})$. We find massless
Goldstone modes that should be candidates for the longitudinal modes of 
massive
gauge bosons, and scalar modes that should be a candidate for Higgs
particles. 

\vspace*{0.5cm}
\noindent
{\bf 4.}\hspace*{0.3cm}
So far, we know that the gap equation (\ref{gap1}) can have the following three 
possible solutions for the quark mass matrix in the quark family space (electrical charges $Q={2\over 3},
-{1\over 3}$)
\begin{equation}
\left(\matrix{m & 0 & 0\cr 0 & m & 0\cr 0 & 0 & m}\right)^{n=36}_Q
(i);\left(\matrix{0 & 0 & 0\cr 0 & m & 0\cr 0 & 0 & m}\right)^{n=16}_Q
(ii);\left(\matrix{0 & 0 & 0\cr 0 & 0 & 0\cr 0 & 0 & m}\right)^{n=4}_Q(iii),
\label{matrix}
\end{equation}
where $n$ stands for the number of Goldstone modes or scaler modes 
associated with each solution. In order to ascertain which solution is physically
realizable, 
we turn to the computation of the ground
state energy. In the one-loop approximation ($O(N_c)$), the effective potential
upon the occurrence of this soft spontaneous symmetry breaking 
is given by
\begin{equation}
V(m, r)={m^2\over G_1}-N_c\tr\int_l\ln
\{ {\gamma_\mu\sin l_\mu\over a}+(m+{r\over a}w(l))\}+\cdot\cdot\cdot,
\label{eff}
\end{equation}
and the difference between the energy
of the symmetric vacuum and broken vacuum $\Delta E_\circ=V(m,r)-V(0,0)$ is
given by 
\begin{equation}
\Delta E_\circ= -{2N^m_g \over a^4} \int_l \sum^{\infty} _{k=1} 
{2N_c \over k+1} 
\left[{(ma + rw(l))^2 \over s^2(l) + (ma + r w(l))^2} \right]^{k+1},
\label{vac1}
\end{equation}
which is obtained from (\ref{eff}) by considering the gap equation
(\ref{gap1},\ref{gap2}). The negative $\Delta E_\circ$ shows that the non-trivial solutions of the gap
equations characterize a chirally asymmetric vacuum that has an energy density
lower than that of the symmetric vacuum. However, it shows that more quark 
families ($N^m_g$) acquire masses the lower ground state energy is, which leads us to
select the first quark mass matrix $(i)$ in eq.({\ref{matrix}). 
A phenomenological disaster occurs, for quarks get equally massive and, 
furthermore, 36 Goldstone modes appear.

On the other hand, noticing that composite bosons
give a positive energy density to such broken ``vacua'' and this positive
contribution certainly increases as the number of composite modes increases, 
we turn to the
computation of the total vacuum energy ($\Delta E$) (vacuum bubble diagrams)
containing both quark 
and composite mode contributions on the basis of gap equations
(\ref{gap1},\ref{gap2}), 
\begin{equation}
\Delta E = -\left[\ln\int_f\exp(-S_{eff}(m,r))-\ln\int_f\exp(-S_{eff}(0,0))
\right],
\label{tol}
\end{equation}
where $S_{eff}(m,r)$ is the effective Wilson action over the ground state.
The details of the calculation are lengthy and will not be reported in this
letter, we just present the result( $\Delta E_\circ$ is $O(N_c)$ and the 
second term $O(N_c^0)$): 
\begin{eqnarray}
\Delta E &\simeq& \Delta E_\circ-(2N^m_g)^2\Big[1-e^{-\Delta E_\circ}\Big]\nonumber\\
&&\cdot\left[-4+\int_l\big(4g_1\tilde\Gamma_s(l)
+{1\over4g_1\tilde\Gamma_s(l)}\big)\!+\!
\int_l\big(4g_1\tilde\Gamma_p(l)+{1\over4g_1\tilde\Gamma_p(l)}
\big)\right],
\label{vac2}
\end{eqnarray}
where
\begin{eqnarray}
\tilde\Gamma_p(l)&=&{a^2I_p(l)A(l^2)\over 4N_c};\nonumber\\
\tilde\Gamma_s(l)&=&{a^2\over 4N_c}\left({4N_c \over a^2} \int_k 
{[ma +rw(k)]^2 \over 
D(k,l)} + I_s(l)A(l^2)\right).
\label{extra}
\end{eqnarray}
Combining positive and negative contributions in eq.~(\ref{vac2})
and putting $ma\simeq 0$, $N_c=3$ and $g_1\simeq g_1^c(r)$ obtained from
eq.~(\ref{gap1}), we find (Fig.(2)) that
the solution $(iii) N^m_g=1$ in eq.(\ref{matrix}) is the energetically 
favoured solution. This shows that, through
this mechanism, only one quark family acquires mass 
and the other quark families remain massless. We thus give the names of top 
and bottom
to this massive quark family. The three Goldstone modes ($\langle \bar t\gamma_5
b\rangle, \langle \bar b\gamma_5
t\rangle$ and ${1\over \sqrt{2}}(\langle \bar t\gamma_5 t\rangle-\langle 
\bar b\gamma_5
b\rangle)$ should become the longitudinal modes of the intermediate gauge bosons.

\vspace*{0.5cm}
\noindent
{\bf 5.}\hspace*{0.3cm}
As has been seen, this research provides evidence and a ``raison d'\^etre''
for the hyerarchy structure of the quark spectrum and thus a
theoretical motivation for the $\bar tt $-condensate model. However, composite
scalar modes disappear from the low-energy spectrum since their masses are
proportional to ${r\over a}$ ($4m^2_s=4m^2+0.8r{m\over a}+0.9{r^2\over a^2}$
obtained from eq.(\ref{scalar})), since in this study, there is no symmetry to
protect their masses from being contributed to by doublers. Whether Higgs
masses are pinned down by gauge interactions or other reasons will be the 
subject of future work. If we turn on gauge fields and couple them to quarks in the
action (\ref{action}), the ``Rome approach'' should be very powerful to establish
the link between heavy quark masses and intermediate gauge bosons ($W^\pm,
Z^\circ$) masses\cite{xueg}. As for splitting the degeneracy of top and bottom
quark masses \cite{xuetb} and the mass generation of other light quark masses,
we suspect that this should eventually occur owing to gauge interactions and
mixing between quark families.

\newpage  \pagestyle{empty} 
\begin{center} \section*{Figure Captions} \end{center}
 
\vspace*{1cm}
 
\noindent {\bf Figure 1}: \hspace*{0.5cm} 
The critical line $g_1^c(r)$, where $m=0$, in terms of $r$.

\noindent {\bf Figure 2}: \hspace*{0.5cm} 
The vacuum energy $\Delta E(r)$ for different massive quark families $N_g^m$.

}

\begin{thebibliography}{99}

\bibitem{nambu1}
Y.~Nambu, in New theories in physics, Proc.\ XI Int.\ Symp.\ on elementary 
particle physics, eds.~Z.~Ajduk, S.~Pokorski and A.~Trautman (World Scientific,
Singapore,~1989).

\bibitem{bar}
W.A.~Bardeen, C.T.~Hill and M.~Linder,{\sl Phys.~Rev.} {\bf D41} (1990) 1647;\\
V.A.~Miranski, M.~Tanabashi and K.~Yamawaki, {\sl Mod.~Phys.~Lett.} {\bf A4}
(1989) 1043; 
{\sl Phys.~Lett.} {\bf B221} (1989) 117;\\
W.J.~Marciano, {\sl Phys.~Rev.~Lett.} {\bf 62} (1989) 2793; {\sl Phys.~Rev.}
{\bf D41} (1990) 219;\\
W.A.~Bardeen, C.T.~Hill and S.~Love, 
{\sl Nucl.~Phys.} {\bf B323} (1989) 493,\\
W.A.Bardeen, S.~T.~Love and V.A.~Miransky, {\sl Phys.\ Rev.} {\bf D42} (1990) 
3514;\\
M.~Lindner, {\sl Int.~Mod.~Phys.} {\bf A8} (1993) 2167.

\bibitem{nambu2}
Y.~Nambu and G.~Jona-Lasinio, {\sl Phys.~Rev.} {\bf 122} (1961) 345.

\bibitem{nogo}
H.B.~Nielsen and M.~Ninomiya, {\sl Nucl.~Phys.} {\bf B185} (1981) 20, {\it
ibid.} {\bf B193} (1981) 173, {\sl Phys.~Lett.} {\bf B105} (1981) 219.

\bibitem{ep}
E.~Eichten and J.~Preskill, {\sl Nucl.~ Phys.} {\bf B268} (1986) 179.

\bibitem{xue91}
G.~Preparata and S.-S.~Xue, {\sl Phys.~Lett.} {\bf B264} (1991) 35;
{\sl Nucl.~Phys.} {\bf B26} (Proc.~Suppl.) (1992) 501;
{\sl Nucl.~Phys.} {\bf B30} (Proc.~Suppl.) (1993) 647.

\bibitem{wheeler}
We also recall that proposals by C.W.~Misner, K.S.~Thorne and J.A.~Wheeler, 
{\it Gravitation\/} (Freeman, San Fransisco, 1973) exist, based on the
violent quantum fluctuations of the metric field at space-time distances of the
order of the Planck-length $a_p\sim10^{-33}$cm, that endow space-time with a
``foam-like" structure of grain-size $a_p$, thus making it equivalent to a
4-dimensional random lattice of lattice constant $a_p$, which we may call a
Planck lattice.

\bibitem{wilson}
K.~Wilson, in {\it New phenomena in subnuclear physics\/} 
(Erice, 1975) 
ed.\ A.~Zichichi (Plenum, New York, 1977).

\bibitem{rome} 
A.~Borrelli, L.~Maiani, G.C.~Rossi, R.~Sisto and M. Testa, {\sl Nucl.~ Phys.}
{\bf B333} (1990) 335; L.~Maiani, G.C.~Rossi, R.~Sisto and M. Testa,
{\sl Phys.~Lett.} {\bf B221} (1989) 360; {\it ibid} {\bf 261} (1991) 479.

\bibitem{smit} L. H. Karsten and J. Smit, {\sl Nucl. Phys.} {\bf B144}
(1978) 536.

\bibitem{xueg}
G.~Preparata and S.-S.~Xue, {\sl Phys.~Lett.} {\bf B329} (1994) 87.

\bibitem{xuetb}
G.~Preparata and S.-S.~Xue, {\sl Phys.~Lett.} {\bf B325} (1994) 161.

\end{thebibliography}
\end{document}